\documentclass[twocolumn,10pt]{revtex4}%
\usepackage[paperwidth=210mm,paperheight=297mm,centering,hmargin=2cm,vmargin=2.5cm]{geometry}
\usepackage{amsfonts}
\usepackage{amsmath}
\usepackage{amssymb}
\usepackage{bm}
\usepackage{graphicx}
\usepackage{color}
\usepackage{soul}
\usepackage{epsfig}%
\usepackage[dvipsnames]{xcolor} 
  \definecolor{bleu_cite}{RGB}{0,0,255}
\usepackage[colorlinks=true,allcolors = black,citecolor=bleu_cite,  ]{hyperref} 
\usepackage{natbib}

\usepackage{siunitx}

\begin{document}
\title{A Microscopic Lattice for Two-dimensional Dipolar Excitons}
%{Universality and Phase Diffusion of a Strongly Interacting  Two-Dimensional Dipolar Gas}

\author{Camille Lagoin$^1$, Stephan Suffit$^{2}$, Mathieu Bernard$^1$, Maxime Vabre$^1$, Kenneth West$^3$, Kirk  Baldwin$^3$, Loren Pfeiffer$^3$ and Fran\c{c}ois Dubin$^{1}$} 
\affiliation{$^1$ Institut des Nanosciences de Paris, CNRS and Sorbonne Universit{\'e}, 4 pl. Jussieu,
75005 Paris, France}
\affiliation{$^2$ Laboratoire de Materiaux et Phenomenes Quantiques, Universite Paris Diderot, Paris}
\affiliation{$^3$ PRISM, Princeton Institute for the Science and Technology of Materials, Princeton Unviversity, Princeton, NJ 08540}

\begin{abstract}
We report a two-dimensional artificial lattice for dipolar excitons confined in a GaAs double quantum well. Exploring the regime of large fillings per lattice site, we verify that the lattice depth competes with the magnitude of exciton repulsive dipolar interactions to control the degree of localisation in the lattice potential. Moreover, we show that dipolar excitons radiate a narrow-band photoluminescence, with a spectral width of a few hundreds of \SI{}{\micro\eV} at \SI{340}{\milli\kelvin}, in both localised and delocalised regimes. This makes our device suitable for explorations of dipolar excitons quasi-condensation in a periodic potential.
\end{abstract}

\maketitle

Electrically polarised GaAs double quantum wells provide a model environment to study cold exciton gases. In these heterostructures lowest energy exciton states are made of a hole localised in one quantum well and Coulomb bound to an electron confined in the other quantum well \cite{Combescot_ROPP}. Such dipolar excitons are long-lived ($\gtrsim$ \SI{100}{\nano\second}), whereas they efficiently thermalise to sub-Kelvin temperatures in this two-dimensional geometry \cite{Oh_Singh,Ivanov_2004}. Dipolar excitons are then possibly studied at thermodynamic equilibrium \cite{Beian_2017,Dang_2020}, and in a homogeneously broadened regime dominated by exciton-exciton interactions \cite{Dang_2020}. Such unique physical properties have led to signatures \cite{Beian_2017,Anankine_2017,Dang_2019,Dang_2020} of excitons quasi-condensation at sub-Kelvin bath temperatures, compatible with a Berezinskii-Kosterlitz-Thouless crossover.

Dipolar excitons are characterised by their permanent electric dipole moment controlled by the separation $L$ between the two quantum well centers. The latter is typically of the order of \SI{10}{\nano\metre} so that the electric dipole moment easily reaches 100 Debye. Exciton-exciton interactions are therefore controlled by a strong repulsive dipolar potential \cite{Ivanov_2010,Schindler_2008,Rapaport_2009}. Furthermore, the excitons dipole moment is by construction aligned perpendicular to the quantum wells. This implies that dipolar excitons have a potential energy controlled by the amplitude of the electric field applied orthogonally to the bilayer. By engineering a spatially inhomogeneous electric field in the plane of a GaAs double quantum well, typically using a set of gate electrodes deposited at the surface of a field-effect device embedding a GaAs bilayer, a rich variety of trapping potentials have been demonstrated \cite{Chen_2006,Schinner_2013,High_2009,Shilo_2013}, as well as devices where excitonic transport is controlled \cite{Gross_2009,Winbow_2011,Kuznetsova_2015,Butov_review_2017}.

In this work, we report on a microscopic square lattice distributing periodically dipolar excitons in the plane of a GaAs double quantum well (DQW). As in earlier studies \cite{Remeika_2011}, the lattice potential is created by a pair of interdigitated semi-transparent electrodes deposited at the surface of a field-effect device embedding a GaAs bilayer. In this report, we show that the interplay between the lattice depth and the strength of repulsive dipolar interactions between excitons controls the degree of exciton localization in the lattice sites. Furthermore, for $n\sim$ \SI{2e10}{\per\square\centi\meter} we show that the photoluminescence spectral width is bound to a few \SI{100}{\micro\eV} at \SI{340}{\milli\kelvin}, for both localised and delocalised regimes. We have previously reported that collective phenomena can then become dominant \cite{Anankine_2017,Dang_2019,Dang_2020}. Studying excitons quasi-condensation in a periodic potential is then experimentally accessible, as discussed in a separate work \cite{Lagoin_2020}.

\begin{figure}[!ht]
  \includegraphics[width=\linewidth]{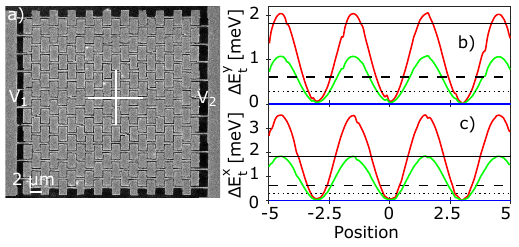} \label{fig:fig1}
  \caption{a) Electron microscope image of the 2D, \SI{3}{\micro\meter} period, square lattice device. One every two electrode row is set to an electrostatic potential $V1$ (left) and the other one to a potential $V2 <V1$ (right), so that the $V2$ ($V1$) electrodes implement traps (barriers) for exciton transport. The white lines in (a) highlight the directions along which we compute the profile of the trapping potential $E_t$, along the vertical axis (b) and along the horizontal axis (c). In (b-c) the profiles are calculated for a potential difference $\Delta V=(V1- V2)$ set to \SI{3}{\volt} (red), \SI{1.5}{\volt} (green) and \SI{0}{\volt} (blue). The horizontal plain/dashed/dotted dark lines mark the PL blueshift energy \SI{0}{\nano\second}, \SI{160}{\nano\second}, \SI{250}{\nano\second} after extinction of the laser pulse loading excitons in the structure.}
\end{figure}

At the heart of our experiments lie two \SI{8}{\nano\metre} wide GaAs quantum wells separated by a \SI{4}{\nano\metre} Al$_{.3}$Ga$_{.7}$As barrier. \textcolor{black}{The heterostructure is embedded in a Al$_{.3}$Ga$_{.7}$As based field-effect device, \SI{150}{\nano\metre} above a $n$-doped GaAs substrate acting as electrical ground, and \SI{1.5}{\micro\metre} below the surface where an array of interdigitated gate electrodes is deposited. With these geometrical factors detrimental electric fields in the plane of the double quantum well are minimized \cite{Rapaport_2005}.} Figure 1.a shows a scanning electron microscope image of the surface electrodes, which realise a square pattern with \SI{3}{\micro\meter} spatial period. By applying on every two row a constant potential $V1$ and on the other ones a potential $V2 <V1$ we imprint a spatially periodic electric field E$_\mathrm{z}$, in the direction perpendicular to the DQW plane. Relying on its interaction with the exciton electric dipole $-eL$, where $-e$ denotes the electron charge while $L$= \SI{12}{\nano\metre}, we control the exciton potential energy that reads $E_t$(\textbf{r})=$(-e.L)\cdot$E$_\mathrm{z}$(\textbf{r}).

We performed finite element simulations to calculate the amplitude of E$_\mathrm{z}$ in the DQW plane as a function of the potential difference $\Delta V$=($V_1$-$V_2$). We then deduced the spatial profiles of the exciton potential energy along the vertical and horizontal directions of the lattice (see Fig.1.b and 1.c respectively). Evaluating the lattice depth $E_t$ for three potential differences, namely $\Delta V=$\SI{3}{\volt} (red), \SI{1.5}{\volt} (green) and \SI{0}{\volt} (blue), we verify that varying $\Delta V$ allows us to tune the confinement profile from a ``flat'' potential with no energy modulation (blue), to a lattice with a depth $\Delta E_t$ up to \SI{3}{\milli\eV}  (red). This behaviour reflects that dipolar excitons are high-field seekers so that they minimize their potential energy in the region where $E_z$ is the strongest. Moreover, in Fig.1 we note that the barrier height is not symmetric along the two axis. Instead, lattice sites have a barrier along the horizontal axis about 50$\%$ stronger than the one along the vertical axis.

\begin{figure}[!ht]
  \includegraphics[width=\linewidth]{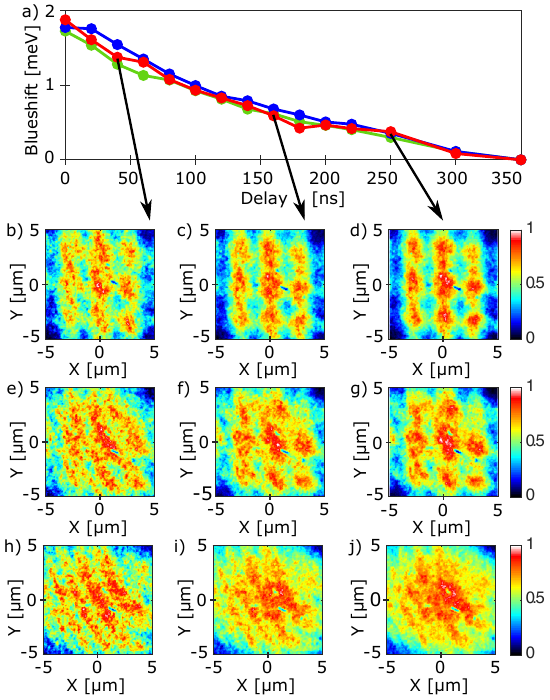}
 \caption{a) Blueshift of the photoluminescence (PL) energy as a function of the delay after the laser pulse extinction for an electrostatic potential difference of \SI{3}{\volt} (red), \SI{1.5}{\volt} (green), \SI{0}{\volt} (blue).
Real images of the normalized PL intensity \SI{40}{\nano\second}, \SI{160}{\nano\second} and \SI{250}{\nano\second} after the laser pulse extinction, for $\Delta V =$ \SI{3}{\volt} (b-d), \SI{1.5}{\volt} (e-g) and \SI{0}{\volt} (h-j). Measurements were all performed at a bath temperature of \SI{340}{\milli\kelvin}.}
\label{fig2}
\end{figure}

In the following we discuss experiments where dipolar excitons were optically injected in the lattice potential. For that, we used a \SI{100}{\nano\second} long laser excitation pulse repeated at \SI{1.5}{\mega\Hz} and tuned at the resonance energy of the direct exciton transition of the two quantum wells, \textcolor{black}{i.e. at 1.574 eV. Thus, we inject electrons and holes directly in both quantum wells, dipolar excitons being formed once carriers have tunnelled towards their minimum energy states located in each layer. Note that our laser excitation is set with an average intensity equal to 1.5 $\mu$W and focussed down to 5 $\mu$m at the center of the lattice device}. At a variable delay to the termination of the loading laser pulse, we analyse the energy of the photoluminescence (PL) radiated by dipolar excitons. Thus, we quantify the exciton density and the strength of their electrostatic confinement. Indeed, the photoluminescence energy scales like $E_t$(\textbf{r})+$u_0n$(\textbf{r}) where the second term reflects the strength of repulsive dipolar interactions between excitons. These lead to a blueshift of the photoluminescence energy, $u_0n$ being of the order of \SI{1}{\milli\eV} when $n$ is about \SI{3e10}{\per\square\centi\meter} \cite{Ivanov_2010,Rapaport_2009,Schindler_2008}. Depending on the delay to the optical loading pulse, the photoluminescence energy maps then the tradeoff between profile of the confining potential and the strength of dipolar repulsions.

In Fig.2.a, we report the dynamics of the photoluminescence energy up to \SI{350}{\nano\second} after termination of the loading laser pulse. For these experiments we evaluate the photoluminescence blueshift, i.e. the difference between the photoluminescence energy at a given delay and its value for the longest delay for which  \textcolor{black}{$u_0n$ is vanishing and the photoluminescence energy is equal to 1.523 eV}. Thus we quantify the dynamics of the exciton population in our electrostatic lattice. For $\Delta V$ ranging from \SI{3}{\volt} to \SI{0}{\volt}, in Fig.2.a we note that the blueshift follows a merely constant decay. This behaviour reveals that the lifetime of dipolar excitons does not depend on $\Delta V$ in our studies, so that at a given delay the average density in the lattice potential is constant regardless the lattice depth. Furthermore, we note that the photoluminescence blueshift right after optical loading does not depend on $\Delta V$. This confirms that the initial exciton density only depends on the intensity of the loading laser pulse, as expected. 

To study the degree of exciton localisation in the lattice potential, we report in Figure 2.b-j real images of the PL measured at three different delays, namely \SI{40}{\nano\second}, \SI{160}{\nano\second}, \SI{250}{\nano\second}, and for $\Delta V$= \SI{3}{\volt} (b-d), \SI{1.5}{\volt} (e-g) and \SI{0}{\volt} (h-j). On each row, $\Delta V$ and therefore the lattice depth are fixed, while on each column the delay is fixed and so is then the mean exciton density. \textcolor{black}{Let us  note that for this entire delay range we expect that dipolar excitons are efficiently thermalised to the bath temperature. Indeed, the strong interaction between excitons and acoustic phonons in GaAs coupled quantum wells ensures that excitons thermalise in at most a few tens of nanoseconds after extinction of the loading laser excitation \cite{Ivanov_2004,Alloing_2011,Alloing_2012}}. 

In Fig. 2 the lowest row displays the regime where dipolar excitons explore a lattice with vanishing barrier height, i.e. a flat potential landscape. Then, at every delay the PL pattern does not reveal any modulation due to the lattice potential. Instead the PL is rather homogeneous spatially, illustrating that dipolar excitons are delocalised in the plane of the double quantum well. We only note a slight curved modulation in the PL images, \textcolor{black}{of around 10 $\%$ amplitude, due to Newton interference fringes created by the intensifier coupled to our charged coupled device camera. Such interference can not be avoided for spectrally narrowband images as the ones we study here. Nevertheless these do not limit our analysis, as shown below.} By contrast, for the first row (Fig.2.b-d) where the lattice depth is the greatest, the PL  exhibits a \SI{3}{\micro\meter} period and square modulation, at every delay studied. This behaviour was expected since in these experiments the lattice depth always exceeds the photoluminescence blueshift (see Fig.1). Also, we note in Fig.2.b (at the termination of the loading laser pulse) that the localisation is stronger along the horizontal axis. Again, this behaviour was expected from our simulations (Fig.1) since along the horizontal axis the lattice depth is the greatest. Increasing the delay, the difference between the horizontal and vertical directions fades away since the PL blueshift decreases.

The photoluminescence images displayed in the middle row of Fig.2 highlight the intermediate regime where the localisation induced by the lattice potential competes with repulsive dipolar interactions between excitons. Indeed, for $\Delta V$= \SI{1.5}{\volt} the lattice depth $\Delta E_t$ ranges from about \SI{1.7}{\milli\eV} in the horizontal axis to \SI{1}{\milli\eV} in the vertical one (Fig.1.b-c). Short after termination of the loading laser pulse, these amplitudes are smaller or of the same order as the photoluminescence blueshift (around \SI{2}{\milli\eV}). Accordingly, Fig.2.e shows that the exciton gas is not localised by the lattice potential at the termination of the loading laser pulse, i.e. when the photoluminescence blueshift is the largest. Increasing the delay to the loading pulse to \SI{160}{\nano\second} (Fig.2.f) the blueshift decreases to about \SI{0.7}{\milli\eV} so that dipolar repulsions no longer suffice to overcome the potential barrier between the lattice sites and the gas becomes then localized in the lattice sites. This localisation is naturally better marked for longer delays for which the photoluminescence blueshift is reduced further (Fig.2.h).

\begin{figure}[!ht]
  \includegraphics[width=.9\linewidth]{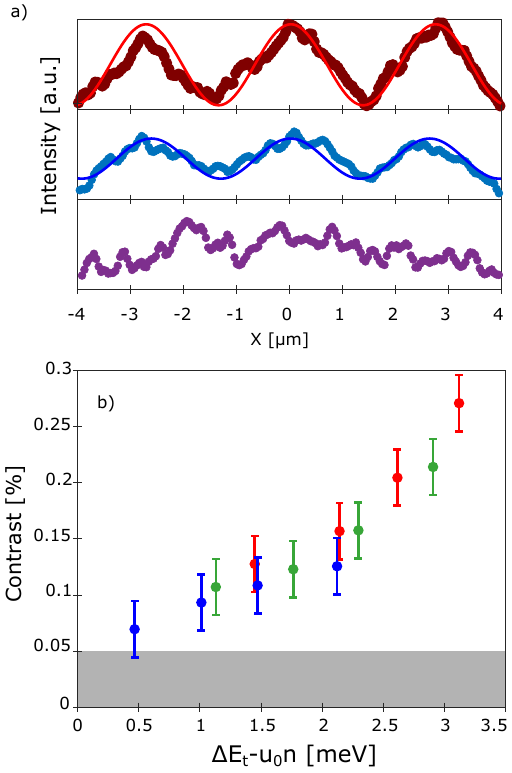}
  \caption{ a) Photoluminescence profiles along the horizontal axis of the lattice measured \SI{160}{\nano\second} after extinction of the loading laser pulse. The top panel is measured for $\Delta V =$ \SI{3}{\volt} (red) and fitted by a sinusoidal function with 21 $\%$ contrast (solid line); The middle panel for $\Delta V =$ \SI{1.5}{\volt} (blue) yields a contrast of about 10 $\%$ while for $\Delta V =$ \SI{0}{\volt} (bottom panel) no intensity modulation is observable. (b) Contrast of the intensity modulation as a function of the effective barrier height ($\Delta E_t-u_0 n$) obtained by varying $\Delta V$ and the delay after the termination of the laser pulse. The points measured for a delay set to \SI{40}{\nano\second} are displayed in blue, for \SI{160}{\nano\second} in green and for \SI{250}{\nano\second} in red. The gray area marks the minimum contrast possibly extracted due to the signal to noise ratio floor for our measurement. Measurements were all performed at a bath temperature of \SI{340}{\milli\kelvin}.}
  \label{fig:fig3}
\end{figure}

For a mean exciton density $n\sim$ \SI{2.5e10}{\per\square\centi\meter}, obtained when the delay to the loading pulse is set to \SI{160}{\nano\second}, the measurements displayed in Fig.2.c, Fig.2.f and Fig.2.i show that we tune the exciton localisation by varying the potential difference between our surface electrodes $\Delta V$. To quantify this degree of control, in Fig.3.a we report the profile of the photoluminescence intensity along the horizontal direction. For the greatest lattice depth (\SI{3.5}{\milli\eV}, Fig.2.c), the upper curve in Fig.3.a reveals a sinusoidal modulation of the photoluminescence intensity, with a \SI{3}{\micro\meter} period and a contrast equal to 21 $\%$. For a moderate lattice depth, $\Delta E_t=$\SI{1.8}{\milli\eV} (Fig.2.f), the middle curve in Fig.3.a exhibits a lower modulation amplitude of around 10 $\%$. Finally, for a ''flat'' potential (lower curve in Fig.3.a) no intensity modulation is observed. 
 
Fig.3.b summarizes the variation of the PL intensity modulation as a function of the effective lattice depth obtained by subtracting $u_0 n$ to the barrier height $\Delta E_t$. Considering this effective lattice  height allows us to directly compare experiments realised for varying exciton density and lattice depth. For clarity measurements performed for the same mean density $n$ are displayed using the same colour, and remarkably we observe that our experimental results all follow a single scaling. This behaviour signals directly that the modulation of the photoluminescence intensity results from the competition between the lattice depth and the strength of repulsive dipolar interactions between excitons. From Fig.3.b we deduce that exciton localisation is possibly detected when the modulation amplitude $\Delta E_t$ is approximately \SI{0.5}{\milli\eV} greater than $u_0 n$. \textcolor{black}{ Finding a threshold here is not completely surprising since our optical resolution, of around \SI{1}{\micro\metre}, has to be taken into account. Indeed, for the parameter range explored in Fig.3.b, Fig.1.b signals that for $(\Delta E_t-u_0 n)$ $\sim$ \SI{0.5}{\milli\eV} the spatial separation between the photoluminescence emitted by nearest lattice sites can not exceed \SI{1.5}{\micro\metre}. As a result, the overlap between the emissions of adjacent lattice sites is significant given our spatial resolution.}

\begin{figure}[!ht]
  \includegraphics[width=\linewidth]{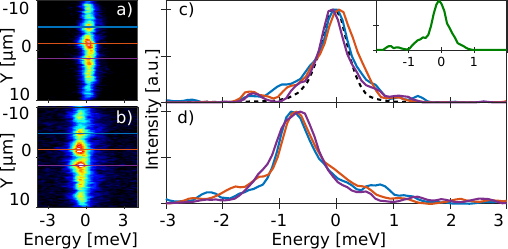}
  \caption{Spatially resolved PL spectra for $\Delta E_t=$ \SI{2}{\milli\eV} and for two delays after extinction of the laser pulse, \SI{140}{\nano\second} (a) and \SI{300}{\nano\second} (b). (c-d) Spectra averaged over \SI{1}{\micro\meter} for three different lattice sites (horizontal lines in (a-b)). All spectra are normalized and centered around the central energy of the PL emitted for a delay set to \SI{140}{\nano\second} (panel (c)). The dashed line in (c) corresponds to the spectral resolution measured with a Hg lamp while the inset displays a spectrum measured for $\Delta E_t=$ \SI{0}{\milli\eV} (flat potential) and the same exciton density as in (c). Measurements were all performed at a bath temperature of \SI{340}{\milli\kelvin}.}
  \label{fig:fig4}
\end{figure}

To assess the performance of the lattice potential and its relevance to explore excitons quasi-condensation in a periodic potential, we finally studied the PL spectrum along the lattice. In previous reports we have shown that quasi-condensation is bound to sub-Kelvin temperatures and to the regime where excitons explore a model electrostatic environment, with a minimum concentration of free carriers \cite{Anankine_2018}. This regime is then signaled by a photoluminescence spectral width lying in the range of a few hundreds of \SI{}{\micro\eV} for $n\sim$ 2-3 10$^{10}$ cm$^{-2}$ \cite{Dang_2020}, otherwise quantum signatures are easily blurred by inhomogeneous broadening \cite{Alloing_2011}.

Figure 4 shows the spatially resolved photoluminescence spectra for $\Delta E_t$ set to \SI{2}{\milli\eV} at two different delays after the extinction of the laser pulse (\SI{140}{\nano\second} for panel (a) corresponding to \SI{0.73}{\milli\eV} blueshift, and \SI{300}{\nano\second} for panel (b) corresponding to \SI{0.08}{\milli\eV} blueshift). Three lattice sites are clearly visible for both delays. The coloured horizontal lines in Fig.4.a and Fig.4.b underline the site centers around which we extract the photoluminescence spectra shown in Fig.4.c and Fig.4.d respectively. Strikingly, for both delays we note that the spectrum exhibits negligible variations between the lattice sites, namely it displays the same lineshape and the same emission energy. This behaviour reveals that our electrostatic lattice is highly regular. Indeed, the spectra displayed in Fig.4.d are measured in the regime where the PL blueshift is negligible compared to the lattice depth. Thus, we deduce that the minimum energy of the lattice sites has negligible variations. On the other hand, from the spectra displayed in Fig.4.c, we conclude that lattice sites are rather uniformly filled, with around 200 excitons per site. \textcolor{black}{ Indeed the three sites analysed here yield a photoluminescence emitted at the same energy, with the same profile that further suggests that thermodynamic equilibrium is reached across the lattice \cite{Andreev_prl} }. Moreover, in Fig.4.c the dashed line displays the spectral resolution for these measurements, obtained using a Hg emission line. It shows that the minimum spectral width possibly detected is around \SI{0.6}{\milli\eV}. Compared to this limit, the spectra at moderate and at dilute densities displayed in Fig.4.c and Fig.4.d have an average full width at half maximum of \SI{640}{\micro\eV} and \SI{850}{\micro\eV} respectively. The purple spectrum in Fig.\ref{fig:fig4}.c, for which we measure a width of \SI{570}{\micro\eV} is even limited by our spectral resolution. A similar limitation is found in the regime where the lattice depth is negligible ($\Delta E_t\sim$ \SI{0}{\milli\eV} in the inset of Fig.4.c). This shows that the PL has a spectral width bound to a few 100 $\mu$eV.

To summarize we have characterised an electrostatic lattice to periodically confine dipolar excitons in the plane of a GaAs double quantum well. We have confirmed that such a confinement potential is efficiently prepared using a set of interdigitated electrodes polarised independently. Furthermore, we have quantified the transition between the regime where optically injected excitons are localised by the lattice potential and the regime where on the contrary they are delocalised. We have shown that this transition is controlled by the competition between the depth of the lattice potential and repulsive dipolar interactions between excitons. Limited by a \SI{1}{\micro\metre} optical resolution, we observe that excitons become localised in the lattice sites when the lattice depth is about 0.5-\SI{1}{\milli\eV} greater than the repulsive dipolar interaction energy. Finally, we have verified that in the lattice potential dipolar excitons radiate a narrow-band photoluminescence, in both localised and delocalised regimes. This underlines the high regularity of the electrostatic potential, as necessary to explore excitons quasi-condensation in a periodic potential. This quasi-condensation is discussed in an independent report \cite{Lagoin_2020}.

\section*{Acknowledgments}
We would like to thank M. Zamorano for her contribution during the early stage of sample development. Our work has been financially supported by the Labex Matisse and by OBELIX from the French Agency for Research (ANR-15-CE30-0020). The work at Princeton University was funded by the Gordon and Betty Moore Foundation through the EPiQS initiative Grant GBMF4420, and by the National Science Foundation MRSEC Grant DMR 1420541.

\end{document}